\documentclass[aps,prd]{revtex4}
\usepackage[colorlinks=true, pdfstartview=FitV, linkcolor=blue, citecolor=red, urlcolor=magenta]{hyperref}
\usepackage{graphicx}
\usepackage{latexsym}
\usepackage{amsmath}
\usepackage{amsfonts}
\usepackage{amssymb}
\usepackage{verbatim}
\begin{document}
\title{Gravitational Dipole Moment in Braneworld Model}

\author{$^{1,2,3}$ Eug\^enio Bastos Maciel}
\email{eugenio.maciel@df.ufcg.edu.br}

\author{$^{1}$M. A. Anacleto}
\email{anacleto@df.ufcg.edu.br}

\author{$^{1}$ E. Passos}
\email{passos@df.ufcg.edu.br}

\affiliation{$^{1}$Unidade Acad\^emica de F\'{\i}sica, Universidade Federal de Campina Grande,\\
Caixa Postal 10071, 58429-900, Campina Grande, Para\'{\i}ba, Brazil.}

\affiliation{$^{2}$Unidade Acad\^emica de Engenharia de Produ\c{c}\~ ao, Universidade Federal de Campina Grande,\\
Caixa Postal 10071, 58540-000, Sum\'e, Para\'{\i}ba, Brazil.}

\affiliation{$^{3}$Departamento de F\'isica, Universidade Estadual da Paraíba,\\
Caixa Postal 781/791, 58429-500, Campina Grande, Para\'{\i}ba, Brazil.}
\begin{abstract}
We investigate the gravitational effects on the relativistic Dirac theory of a system such as a Hydrogen atom in the braneworld scenario. A gravitational dipole moment like contribution, arises in the nonrelativistic Hamiltonian of the system through an exact Foldy-Wouthuysen transformation. 
This term violates the equivalence principle for the weak interaction that is restored in the average over the spins. Furthermore, it feels the effects of the extra dimensions, so that in a Universe with two additional spatial dimensions its energy contribution is amplified by an order of $\sim 10^{16}\ \text{eV}$ concerning to the energy of this term for ordinary space. The compactification radius for a Universe with two extra dimensions is within the experimental limits, where deviations from the inverse square law are being tested. This suggests that the energy value for the gravitational dipole term in this scenario may lead us to search for traces of extra dimensions in atomic spectroscopy, as well as experimental constraints for these dimensions.
\end{abstract}
\pacs{11.15.-q, 11.10.Kk} 
\maketitle

\pretolerance10000
\section{Introduction}
It is known that gravitational interaction is very weak when compared to other fundamental interactions, which justifies the fact that we neglect its effects in the atomic domain. However, in recent decades, this scenario has been taking a new direction. Due to technological advances, especially in the interferometry technique, it has become feasible to investigate the effects of gravitational effects in atomic spectroscopy \cite{Colella:1975dq,Bonse:1983dda}. Therefore, it is expected that in the near future, the investigation of gravitational effects on an atomic scale will become an important line of research in experimental physics, even with great challenges. On the other hand, theoretical physics always provides viable results for testing the influence of the gravitational field on quantum systems \cite{kobzarev1962,Leitner:1964tt,Peres:1977yh,morgan1962,Mashhoon:2000jq}. As an example, Peres' \textit{ad hoc} model \cite{Peres:1977yh} in which the Dirac non-relativistic Hamiltonian should contain a spin gravitational dipole moment like term $\pm k\hbar c^{-1}\vec{\sigma}\cdot\vec{g}$ where $\vec{g}$ is the gravitational acceleration vector and $k$ is the dimensionless coupling constant. The constants $c=3\times10^{8}\ \text{m/s}$ and $\hbar=6,582119\times10^{-16}\text{eV}\cdot{\rm s}$ are the speed of light and the Planck constant respectively. This term imposed on the Dirac theory preserves CP symmetry but not C and P in separate forms and must violate the equivalence principle of weak interaction, which should be restored at a macroscopic level. 

Namely, spin gravitational dipole moment like contribution, arises naturally in Dirac theory when the non-relativistic limit of the Dirac equation is obtained using an exact Foldy-Wouthuysen transformation \cite{Eriksen:1958zz}. In this context, this term arises in two situations, one for a particle in an accelerated frame of reference and the other for a particle in a gravitational field described by a spherically symmetric system such as the Schwarzschild metric, for example \cite{Obukhov:2000ih}. Another important fact is that the coupling constant assumes a fixed value of $1/2$. Comparison of these theories with experimental data imposes very weak constraints on the coupling constant, since the value $\hbar g/c=2.153\times10^{-23}\ \text{eV}$ is very small.

Motivated by these theoretical perspectives, an alternative to investigating the effects of the gravitational field on atomic spectroscopy would be through theories of extra dimensions in the brane scenario \cite{Arkani-Hamed:1998jmv,Randall:1999ee,Randall:1999vf}. In these models, the Universe is a submanifold embedded in a larger dimensional space known as ambient space or supplementary space where matter and fields are trapped in a three-dimensional space (3-brane) in a confined state where only gravity can propagate throughout the ambient space because it will be the only one spread out in all directions. Thus, the gravitational field could feel the direct effects of the extra dimensions on a length scale larger than the scale where the other fields are located. However, at distances smaller than the size of the extra dimensions, these models predict that the gravitational force should be amplified when compared to the three-dimensional Newtonian gravitational force by a factor that depends on the number of extra dimensions.

Given that tests for the influence of gravitational interaction at short distances are becoming increasingly evident in the field of experimental physics, many researchers have been developing a series of works over the years whose objective, in addition to investigating gravitational effects at the atomic level, of  seeking spectroscopic links for the extra dimensions via the brane scenario\cite{Kapner:2006si,murata2015review,cullen1999sn,frost2009phenomenology,giudice1999quantum,cms2012search,lemos2019s,dahia2016proton,dahia2016constraints,dahia2018rydberg}. In this way, such theories allow the formulation of models that are a priori phenomenological viable, since these models predict the length of extra dimensions much larger than the Planck length $l_{P}=10^{-35}\,m $ which is the length of the radius $\text{R}$ of the extra dimension proposed by the Kaluza-Klein theory \cite{kaluza1921,klein1987}. For example, when testing the inverse square law, the compactification radius must be smaller than $44\mu\text{m}$ \cite{murata2015review,cullen1999sn}. This fact becomes a strict consideration when the ambient space has only one extra dimension. However, if the supplementary space admits more extra dimensions, experimental limits in astrophysics and colliders suggest that this length is larger than the Planck length by at least $15$ orders of magnitude (see this argument in Refs. \cite{frost2009phenomenology,giudice1999quantum,cms2012search}).

In this work, we investigate the influence of extra dimensions on the non-relativistic Dirac Hamiltonian for space with curvature in the brane scenario. We are considering the physical system as the Hydrogen atom. The geometry of this system is of the Reissner-Nordström type, since the gravitational field has two contributions: the gravitational field produced by the mass of the atomic nucleus and the gravitational field produced by the electromagnetic energy that is scattered in space due to electric charge. The gravitational dipole moment term is obtained for the gravitational field produced by the nucleus, with a coupling constant corrected by a factor proportional to the number of extra dimensions. In addition, the gravitational field is also affected by the extra dimensions, amplifying its value when we consider the system in ordinary four-dimensional space-time extra dimensions, which suggests that this term is now energetically larger when compared to the previous ones, which makes feasibility possible. When we consider the gravitational field produced by the electromagnetic energy that spreads through the system, the gravitational dipole moment term appears only as a relativistic correction of the order $1/c^{5}$. This is because this gravitational field due to electromagnetic energy is proportional to $1/c^{4}$. Furthermore, this term depends on the internal structure of the brane, so we neglect its effect on the total Hamiltonian of the system.

This paper is organized as follows. In Sec.~\ref{form}, the formalism of the relativistic Dirac theory for curved spacetime will be explained. The exact Foldy-Wouthuysen transformation for the Hamiltonian will be obtained, which will be used to obtain the non-relativistic limit of the theory for a spherically symmetric system. In Sec.~\ref{asbs}, the atomic system in the brane scenario will be presented. The system metric has the characteristics of the Reissner-Nordström metric, with two contributions to the gravitational field: the field resulting from the mass of the nucleus and the electromagnetic energy of the system. In Sec.~\ref{sgb}, the non-relativistic limit of the Dirac Hamiltonian will be obtained in the brane scenario where the gravitational dipole moment term significantly affects the total energy of the system. Finally, in Sec.~\ref{conc}, the concluding remarks will be presented.
\section{Formalism}
\label{form}
In this section, we describe the Dirac equation in curved space and through which
the nonrelativistic Hamiltonian associated with this system is obtained using the Foldy-Wouthuysen (FW) transformation for Dirac spinor.

\subsection{The Dirac Equation in Curved Space }

In the context of General Relativity Theory, spacetime is defined as a $4-$dimensional Riemannian manifold with a Lorentzian local metric where local Lorentzian spinorial structure  cannot constitute a representation of the general transformations on the manifold \cite{wald1994}. So we make use of a quantity that connects the local flat space with the general structure. This quantity is the basis of tetrads $e^{\hat{a}}_{\mu}$ it is possible to define matrices of Dirac on the manifold
\begin{equation}\label{DE1}
\gamma^{\mu}=e^{\mu}_{\hat{a}}\gamma^{\hat{a}}.
\end{equation}
This definition leads us directly to a generalization of the Clifford algebra for any metric $g^{\mu\nu}$ in curved spacetime.
\begin{equation}\label{DE2}
\gamma^{\mu}\gamma^{\nu}+\gamma^{\nu}\gamma^{\mu}=2g^{\mu\nu}\mathbb{I}.
\end{equation}
Where do we get the relation with the flat metric for Minkowski spacetime $\eta^{\hat{a}\hat{b}}$ is
\begin{equation}
g^{\mu\nu}=e^{\mu}_{\hat{a}}e^{\nu}_{\hat{b}}\eta_{\hat{a}\hat{b}}.
\end{equation}
Here the Latin term `` $\hat{a}$ '' denotes the coordinates of flat spacetime and the Greek indices denote the curved spacetime coordinates. The matrices under the transformation of the Dirac matrices (\ref{DE1}) transform as
\begin{equation}\label{DE3}
\left(\gamma^{\mu}\right)^{\hat{a}}_{\hat{b}}=\left(S\gamma^{\mu}S^{-1}\right)^{\hat{a}}_{\hat{b}},
\end{equation}
this transformation is generated by giving the basis of tetrads, a local Lorentz transformation.
\begin{equation}\label{DE4}
\left(L^{-1}\right)^{\hat{a}}_{\hat{b}}\gamma^{\hat{b}}=S\gamma^{\hat{a}}S^{-1}.
\end{equation}
The condition above suggests that spinor $\bar{\psi}=\psi^{\dagger}\gamma^{\hat{0}}$ transforms as a conjugate spinor. Where $\psi^{\dagger}$ is a conjugate Hermitian of $\psi$ and $\gamma^{\hat{0}}$ is the Dirac matrix constant.

The local group is a point-dependent quantity, so transporting an object to each point on the manifold must contain an affinity term so that the transported object still transforms over the local group \cite{parker2009}. In summary, the fact that the matrices of the transformation group are functions of the point, the derivative of the spinor does not transform like a spinor. In this sense, it is convenient to replace the ordinary derivative by the covariant derivative $\partial_{\mu}\rightarrow\nabla_{\mu}=\partial_\mu+\Gamma_{\mu}$ where the quantity $\Gamma_{\mu}$ is the spinorial connection
\begin{equation}\label{DE9}
\Gamma_{\mu}=-\frac{i}{4}\sigma^{\hat{a}\hat{b}}e^{\nu}_{\hat{a}}\nabla_{\mu}e_{\hat{b}\nu}.
\end{equation}
With $\sigma^{\hat{a}\hat{b}}=\frac{i}{2}[\gamma^{\hat{a}},\gamma^{\hat{b}}]$ being a representation for the Lie algebra in spinor spaces. The covariant derivative of the tetrad field is given in terms of the Christoffel symbols specified by the geometry of the system.
\begin{equation}
\nabla_{\mu}e_{\hat{b}\nu}=\partial_{\mu}e_{\hat{b}\nu}-\Gamma^{\lambda}_{\nu\mu}e_{\hat{b}\lambda}.
\end{equation}
Under Lorentz transformations, the spinorial connection transforms as follows \cite{weinberg1972}
\begin{equation}\label{DE10}
\Gamma_{\mu}\rightarrow S(L)\Gamma_{\mu}S^{-1}(L)-[\partial_{\mu}S(L)]S^{-1}(L).
\end{equation}
Thus, the Dirac equation for curved space is written as the following:
\begin{equation}\label{DE11}
\big(i\hbar\gamma^{\mu}\big(\partial_{\mu}+\Gamma_{\mu}\big)-mc\big)\psi=0.
\end{equation}
This above expression will be explored in a spherical coordinate system.

\subsection{The Spherically Symmetrical System}

One of the simplest applications of the Dirac equation in curved space is to investigate the behavior of a particle in a spherically symmetrical gravitational field.  In a simplified way, the line element of a spherically symmetric space in isotropic coordinates is given by
\begin{equation}\label{SSS1}
ds^{2}=-v^{2}c^{2}dt^{2}+w^{2}\delta_{ij}dx^{i}dx^{j},
\end{equation}
where $w$ and $v$ are scalar functions with dependence on radial coordinates, $r$ ( many
particular cases described by metric (\ref{SSS1}) are widely investigated \cite{fischbach1981,hehl1990,cai1991,varju1998}). Generally, the non-null Christoffel symbols for metrics (\ref{SSS1}) are reduces to
\begin{equation}\label{SSS2}
\Gamma^{0}_{i0}=\frac{1}{2v^{2}}\partial_{i}(v^{2})\quad\mbox{,}\quad
\Gamma^{i}_{00}=\frac{1}{2w^{2}}\partial_{i}(v^{2})\quad\mbox{,}\quad
\Gamma^{i}_{kl}=\frac{1}{2v^{2}}\Big(\partial_{l}\delta_{ik}+\partial_{k}\delta_{il}-\partial\delta_{kl}\Big)(w^{2}).
\end{equation}
The tetrad fields are $e_{\hat{0}0}=v$ and $e_{i\hat{j}}=\delta_{i\hat{j}}w$ and the spinorial connection terms are
\begin{equation}\label{SSS3}
\Gamma_{0}=-\frac{1}{4vw}\vec{\alpha}\cdot\vec{\nabla}(v^{2})\quad\mbox{,}\quad
\Gamma_{i}=-\frac{i}{4v}\Big(\vec{\Sigma}\times\vec{\nabla}(v^{2})\Big).
\end{equation}
We defined the matrices into a Dirac standard representation:
\begin{equation}\label{SSS4}
\vec{\alpha}= \hat{\beta}\vec{\gamma}=\left(\begin{array}{ll}
0 & \vec{\sigma} \\
\vec{\sigma} & 0
\end{array}\right),\;\;
\vec{\Sigma}=\left(\begin{array}{ll}
\vec{\sigma} & 0 \\
0 & \vec{\sigma}
\end{array}\right),
\end{equation} 
where $\Vec{\sigma}=(\sigma_{1}, \sigma_{2}, \sigma_{3})$, $\sigma_{i}\; (i=1, 2, 3)$ 
 are the $2\times 2$ Pauli matrices  \cite{dirac1928}. The  Hamiltonian operator obtained from the Dirac equation (\ref{DE11}) is given as
\begin{equation}\label{SSS5}
\mathcal{H}=G\left(r\right)\beta mc^{2}+\beta mc^{2}-i\hbar c\frac{1}{4wv}\vec{\alpha}\cdot\vec{\nabla}(v^{2})-i\hbar c\frac{v}{2w^{3}}\vec{\alpha}\cdot\vec{\nabla}(w^{2})-i\hbar c\frac{v}{w}\vec{\alpha}\cdot\vec{\nabla},
\end{equation}
where $G(r)=v-1$. The Hamiltonian above must be Hermitian. Indeed, it must have real eigenvalues to allow for the physical interpretation of its results. Thus, the following condition must be satisfied 
\begin{equation}\label{C1}
\langle \mathcal{H}\rangle=\int d^3x \sqrt{g} \phi^{\dagger}(x) \mathcal{H} \psi(x)=\left\langle \mathcal{H}^{\dagger}\right\rangle,
\end{equation}
where  $g$  is the determinant of the spatial part of the metric  $g_{ij}$. The quantities  $\phi(x)$  and  $\psi(x)$  represent possible particle states. It is more convenient to perform the integral using a flat-space measure. In this way, it is necessary to suppress the  $\sqrt{g}$  factor. This is achieved by performing a transformation on the wave function  $\bar{\psi} = \Theta \mathcal{H} \psi$, where  $\Theta$  is a unitary operator such that  $\Theta \Theta^{-1} = 1$. Consequently, the new Hamiltonian  $\bar{\mathcal{H}} = \Theta \mathcal{H} \Theta^{-1}$  satisfies the condition (\ref{C1})
\begin{equation}\label{C2}
\langle \mathcal{\bar{H}}\rangle=\int d^3 x \Theta^2 \phi^{\dagger}(x) \mathcal{H} \psi(x).
\end{equation}
Comparing $(\ref{C1})$ with $(\ref{C2})$ have the relation $\Theta^{2}=\sqrt{g}$. Looking at the metric (\ref{SSS1}), we have that $\Theta=w^{3/2}$, in this sense $\mathcal{H}^{\prime}=w^{3/2}\mathcal{H}w^{-3/2}$ then 
\begin{equation}\label{SSS6}
\mathcal{H^{\prime}}=\beta m c^2 v+\frac{c}{2}[(\vec{\alpha} \cdot \vec{p}) \mathcal{F}+\mathcal{F}(\vec{\alpha} \cdot \vec{p})],
\end{equation}
This Hamiltonian above is Hermitian. Being $\vec{p}=-i\hbar\vec{\nabla}$ and $\mathcal{F}=v/w$. To simplify our notation, from now on we will abandon the bar.

\subsection{The Nonrelativistic Hamiltonian }
In this point, we apply an Foldy-Wouthuysen transformation to derive the nonrelativistic Hamiltonian associate to Eq.(\ref{SSS6}) (the Dirac Hamiltonian). In general the technique decouples positive and negative energy states in which the Dirac Hamiltonian takes the form \cite{foldy1950}:
\begin{equation}\label{FW1}
\mathcal{H}=\beta mc^{2}+\mathcal{O}+\mathcal{E},
\end{equation}
where $\mathcal{O}$ is the odd part and $\mathcal{E}$ is the even part. The odd part carries the Dirac matrices $\Vec{\alpha}$ that mix negative energy states with positive energy, while the even part carries only quantities that do not mix these states, such as the $\beta$ matrix and the identity matrix, for example. Taking equation (\ref{SSS5}) as an example, we have the odd part as 
\begin{equation}\label{C3}
    \mathcal{O}=i\hbar c\alpha\cdot\Big(\frac{1}{4wv}\vec{\nabla}(v^{2})-\frac{v}{2w^{2}}\vec{\nabla}(w^{2})-\frac{v}{w}\vec{\nabla}\Big),
\end{equation}
and the even part as
\begin{equation}
    \mathcal{E}=G(r)\beta mc^{2}.
\end{equation}

The nonrelativistic limit is obtained through an approximate scheme where the odd parts are removed in order by powers of $(1/mc^{2})$
\begin{equation}\label{FW2}
\mathcal{H}=\beta\left[m c^2+\frac{\mathcal{O}^2}{2 m c^2}-\frac{\mathcal{O}^4}{8 m^3 c^6}\right]+\mathcal{E}-\frac{1}{8 m^2 c^4}[\mathcal{O},[\mathcal{O}, \mathcal{E}]]+\cdots .
\end{equation}
It is important to highlight that this method is widely applied in cases of coupling with the electromagnetic field to study gravitational effects, requiring some observations. Roughly speaking, when coupled with the electromagnetic field, parts $\mathcal{O}$ and $\mathcal{E}$ do not depend on the mass $m$ but on the electrical charge $e$ and this justifies the fact that this expansion is taken as a standard \cite{bjorken1964}. So, when we investigate gravitational systems, the $\mathcal{E}$ part necessarily depends on the mass $m$. In this way, although in the first approximation the odd term $\mathcal{O}$ is removed, a new term is produced. The last term, in (\ref{FW2}), carries a term of order $m^{0}$ and this fact occurs in each step of the approximate scheme so that all even terms will always be of the same order in $1/m$ at odd terms are removed in the expansion. This fact leaves the expansion scheme problematic. 

However, in this article, we will use another method to decouple the negative and positive energy states. This method was first proposed by Erick Eriksen and is based on an exact transformation in the Dirac Hamiltonian through a unitary operator $U$ that connects the Dirac representation to the Foldy-Wouthuysen representation \cite{Eriksen:1958zz}. For definition the unitary operator $U$ that connects the Dirac representation with the FW representation so that
\begin{equation}\label{FW3}
U \hat{\Lambda} U^{\dagger}=\beta,
\end{equation}
where $\hat{\Lambda}=\mathcal{H} / \sqrt{\mathcal{H}^2}$ is the Pauli energy operator \cite{pauli1933}. This operator is Hermitian, unitary and idempotent, such that $\hat{\Lambda}^2=\hat{\Lambda}^{\dagger} \hat{\Lambda}=1$. An important fact is that the Hamiltonian (\ref{SSS6}) admits an anticommuting involution operator defined by
\begin{equation}\label{FW4}
J:=i \gamma_5 \beta.
\end{equation}
It is possible to clearly observe that this operator is Hemitian and unitary and anticommutes with the Hamiltonian (\ref{SSS6}) and with the matrix $\beta$.
\begin{equation}\label{FW5}
J \hat{\mathcal{H}}+\hat{\mathcal{H}} J=0, \quad J \beta+\beta J=0 .
\end{equation}
Therefore, we can conclude that the operator that provides the exact FW transformation is written in terms of $U=U_{2}U_{1}$ such that
\begin{equation}\label{FW6}
U_1=\frac{1}{\sqrt{2}}(1+J \hat{\Lambda}), \quad U_2=\frac{1}{\sqrt{2}}(1+\beta J).
\end{equation}
Indeed, we find that $U_1 \hat{\Lambda} U_1^{\dagger}=J$ and $U_2 J U_2^{\dagger}=\beta$. This relationship (\ref{FW5}) leads us directly to the Hamiltonian $\mathcal{H}^{FW}=U \mathcal{H} U^{\dagger}$ in the FW representation \cite{Obukhov:2000ih}
\begin{equation}\label{FW7}
\mathcal{H}=[\sqrt{\mathcal{H}^2}] \beta+\left\{\sqrt{\mathcal{H}^2}\right\} J.
\end{equation}

The Hamiltonian in (\ref{FW7}) is exact and deserves two comments. Firstly, it clearly even in this way and not mix positive and negative energy states. To conclude this statement, we must remember that, in general, the even and odd parts of any operator always have the following form: $[Q]:=\frac{1}{2}(Q+\beta Q \beta)$ and $\{Q\}:=\frac{1}{2}(Q-\beta Q \beta)$ respectively. Second, in general terms, the square of the Hamiltonian is also always a pair operator, which leads us to conclude that the second term of (\ref{FW7}) is always absent. However, this fact is not valid in our system since the square of the Hamiltonian
\begin{equation}\label{FW8}
\mathcal{H}^2=m^2 c^4 V^2+\mathcal{F} c^2 p^2 \mathcal{F} +\frac{\hbar^2 c^2}{2} \mathcal{F}(\vec{\nabla} \cdot \vec{f})-\frac{\hbar^2 c^2}{4} \vec{f}^2+\hbar c^2 \mathcal{F} \vec{\Sigma} \cdot([\vec{f} \times \vec{p}]+J m c \vec{\phi}),
\end{equation}
contains an odd term, the last term of (\ref{FW8}). By definition 
\begin{equation}\label{FW9}
\vec{\phi}:=\vec{\nabla} v, \quad \vec{f}:=\vec{\nabla} \mathcal{F}.
\end{equation}
In general, we consider nonrelativistic wave functions such that all interactions are assumed to be perturbations. The almost relativistic approximation is found when we expand the Hamiltonian (\ref{FW8}) and consider the term $\beta mc^{2}$ the dominant term, so 
\begin{eqnarray}
\label{FW10}
 \sqrt{\hat{\mathcal{H}}^2} &\approx& m c^2 V+\frac{1}{4 m}\left(W^{-1} p^2 \mathcal{F}+\mathcal{F} p^2 W^{-1}\right) +\frac{\hbar^2}{4 m W}(\vec{\nabla} \cdot \vec{f})-\frac{\hbar^2}{8 m V} \vec{f}^2\nonumber\\
&& +\frac{\hbar}{4 m} \vec{\Sigma} \cdot\left(W^{-1}[\vec{f} \times \vec{p}]+[\vec{f} \times \vec{p}] W^{-1}\right. 
 \left.+2J W^{-1} m c \vec{\phi}\right).
\end{eqnarray}
The above expression is the nonrelativistic quantum Hamiltonian for four-component fermions. Namely, the first two terms describe the usual effects already measured experimentally on spinless particles (see Ref. \cite{Colella:1975dq}). The third contribution is the “gravitational Darwin” term which admits a physical interpretation similar to the usual electromagnetic Darwin term, reflecting the zitterbewegung fluctuation of the fermion’s position with the mean square,i.e.,  $\langle(\delta r)^{2}\rangle \sim \hbar/(m c)^{2}$. The first term of the second line gives rise to the gravitational spin-orbit coupling and the last term provides the new term for the interaction of the spin with the gravitational field, the gravitational dipole moment \cite{Peres:1977yh,Obukhov:2000ih,kobzarev1962}. 

\section{Atomic System In Braneworld Scenario}
\label{asbs}
The proposal of the ADD model is to solve the hierarchy problem, which arises from the large discrepancy between the values of the Planck scale and the electroweak scale \cite{Arkani-Hamed:1998jmv}. In this model, our Universe is considered a 3-brane embedded in a bulk space with  $\delta$ extra dimensions with the topology of a torus $T^{\delta}$.
In addition, it is assumed that the energy of the brane does not curve space for long distances when compared to the length scale where the fields are found confined. In this way, it is said that the matter that is confined to the brane is governed by the Einstein-Hilbert action
\begin{equation}\label{TT1}
S_G=\frac{c^3}{16 \pi G_d} \int d^4 x d^\delta z \sqrt{-\hat{g}} \hat{\mathcal{R}}.
\end{equation}
Here $\hat{\mathcal{R}}$ is the scalar of curvature in ambient space and $\hat{g}$ is the determinant of the signed metric $(-,+,\dots,+)$. The constant $G_{D}=G(2\pi R)^{\delta}$ is the gravitation constant for ambient space written in terms of the Newtonian constant $G$. The constant $R$  is the compactification radius, whose relationship with the number of extra dimensions  $\delta$  is established according to the relation. See Ref. \cite{Arkani-Hamed:1998jmv} 
\begin{equation}\label{TT11}
R \simeq 10^{32 / \delta-19} \mathrm{~m}.
\end{equation}

To obtain a correct Newtonian limit, some mechanism is necessary that provides stabilization of the volume of ambient space for large distances \cite{arkani2001,antoniadis2003}. The coordinates describe the transverse and parallel directions with respect to the brane. In the atomic domain where our system is defined, the metric is $g_{AB}=\eta_{AB}+h_{AB}$ where $g_{AB}$ is the metric in Minkoswski spacetime and $g_{AB}$ is a perturbation of the order of $Gm$. For a coordinate system where the gauge
\begin{equation}\label{TT2}
\partial_A\left(h^{A B}-\frac{1}{2} \eta^{A B} h_C^C\right)=0,
\end{equation}
is satisfied, the linearized Einstein actions obtained from the extremization of (\ref{TT1}) leads us to the equations
\begin{equation}\label{TT31}
\square h_{A B}=-\frac{16 \pi G_d}{c^4} \bar{T}_{A B},
\end{equation}
In the our notation, $\square$ is the d'Alembertian operator and $\bar{T}_{AB}=\left(T_{AB}-(\delta+2)^{-1}\eta^{AB}T^{C}_{C}\right)$ defined in terms of the source tensor $T_{AB}$. Considering the topology $\mathbb{R}^3 \times T^\delta$, the solution of equation (\ref{TT31}) is
\begin{equation}\label{TT3}
h_{A B}(\vec{X})=\frac{16 \pi G_d \Gamma\left(\frac{\delta+3}{2}\right)}{(\delta+1) 2 \pi^{(\delta+3) / 2} c^4} \sum_i \times\left(\int \frac{\bar{T}_{A B}\left(\vec{X}^{\prime}\right)}{\left|\vec{X}-\left(\vec{X}^{\prime}+\vec{K}_i\right)\right|^{1+\delta}} d^{3+\delta} X^{\prime}\right),
\end{equation}
where $\vec{X}=(\vec{x},\vec{z})$ and $\Vec{K}_{i}=2\pi R(0,0,0,k_{1}),\dots,k_{\delta}$ with $k_{i}$ is an integer number. Here the $\Vec{K}_{i}$ vectors are considered as mirror images of the source induced by the topology $T^{\delta}$ of space. In this way, the solutions in relation to the $z$ coordinate become periodic. In this way, it is expected that the Green function recovers the four-dimensional behavior for long distances $|\vec{x}| \gg R$. However, if we are considering small distances 
$\left(\left|\vec{x}-\vec{x}^{\prime}\right|<R\right)$, the Green function is dominated by the first term in the series (\ref{TT3}).
\subsection{The Gravitational Field}

We are interested in investigating the effects of extra dimensions on the gravitational field for short distances where the atomic domain is valid in the braneworld scenario, so we will only take the first term of (\ref{TT3}). It is worth highlighting at this point that for the theory to be phenomenologically viable we must have some mechanism that ensures stabilization for the supplementary space. This procedure is usually adopted in the production of black holes by colliders \cite{giddings2002,dimopoulos2001,argyres1998}. The energy moment tensor for the fields that are confined to the brane has the form
\begin{equation}\label{TT4}
T_{A B}(x, z)=\eta_A^\mu \eta_B^v T_{\mu \nu}(x) f(z),
\end{equation}
where $T_{\mu\nu}(x)$ is the ordinary energy-tensor momentum for the four-dimensional fields in the brane and $f(z)$ is a concentrated function around the brane that must have the appearance of a delta-like function in the thin brane limit. Furthermore, this function describes the confinement of the fields on the brane.

Here the atomic nucleus is the source of the gravitational field, thus a Reissner-Nordström type geometry is expected for our system due to the contribution of electromagnetic energy scattered in space \cite{dahia2018rydberg}. Consequently, it is reasonable to assume that the energy moment tensor has two contributions: $T_{\mu \nu}=T_{\mu \nu}^{(0)}+T_{\mu \nu}^{(E M)}$ with $T_{\mu \nu}^{(0)}$. The tensor $T_{\mu \nu}^{(0)}$ describes the energy concentrated in the atomic nucleus and $T_{\mu \nu}^{(E M)}$ is the moment energy tensor of the electromagnetic energy that is spread out in space, defined respectively as
\begin{equation}\label{TT5}
T_{\mu \nu}^{(0)}=c^{2}\rho\eta^{0}_{\mu}\eta^{0}_{\nu}\quad\mbox{,}\quad
T_{\mu \nu}^{(E M)}=\epsilon_{0}c^{2}\Big(F_{\mu\lambda}F^{\lambda}_{\nu}-\frac{1}{4}F_{\alpha\beta}F^{\alpha\beta}\Big),
\end{equation}
where $\rho$ is the source matter density and $F_{\mu\nu}$ is the electromagnetic tensor with $\epsilon_{0}$ being the electrical permittivity in vacuum. In the first approach, we are neglecting the effects of the magnetic field produced by the proton dipole. Considering these statements, the metrics of our system are described by
\begin{equation}\label{TT6}
\begin{aligned}
d s^2= & -\left(1+\frac{2}{c^2} \varphi_s+\frac{2(2+\delta)}{c^2(1+\delta)} \chi_s\right)\left(d x^0\right)^2 \\
& +\left(1-\frac{2}{c^2(1+\delta)} \varphi_s\right)\left[\left(1+\frac{2(2+\delta)}{c^2(1+\delta)} \lambda_{1, s}\right) d r^2\right. \\
& \left.+\left(1+\frac{2(2+\delta)}{c^2(1+\delta)} \lambda_{2, s}\right) r^2\left(d \theta^2+\sin ^2 \theta d \phi^2\right)\right] \\
& +\left(1-\frac{2}{c^2(1+\delta)} \varphi_s\right) d \vec{z}^2,
\end{aligned}
\end{equation}
the coordinates $r,\theta,\phi$ are the spherical coordinates associated with ``almost Cartesian'' coordinates $(x^{1},x^{2},x^{3})$ and $x^{0}=cdt$. Let’s analyze the metric (\ref{TT6}). The $\varphi_{s}$ function plays the role of the gravitational potential generated by the nucleus and is valid for short distances:
\begin{equation}\label{TT7}
\varphi_s(\vec{X})=-\tilde{G}_D \int \frac{\rho\left(\vec{x}^{\prime}\right) f_m(z)}{\left|\vec{X}-\vec{X}^{\prime}\right|^{1+\delta}} d^{3+\delta} X^{\prime},
\end{equation}
with 
\begin{equation}\label{CG}
\tilde{G}_{D}=\frac{4G_{D}\Gamma\left(\frac{3+\delta}{2}\right)}{(2+\delta)\pi^{(1+\delta)/2}}. 
\end{equation}
The function $\chi_{s}$ is the gravitational potential due to the energy of the electromagnetic field $u=\epsilon_{0} E^{2}/2$ created by the electrical storm
\begin{equation}\label{TT8}
\chi_s(\vec{X})=-\frac{\hat{G}_D}{c^2} \int \frac{u\left(\vec{x}^{\prime}\right) f_e(z)}{\left|\vec{X}-\vec{X}^{\prime}\right|^{1+\delta}} d^{3+\delta} X^{\prime}.
\end{equation}
The functions $\lambda_{2,s}$ and $\lambda_{1,s}$ have their origin in the components of the energy moment tensor for the electromagnetic field and are defined by
\begin{eqnarray}
\label{TT9}
&&\lambda_{2, s}=  -\chi_s-\pi \frac{\hat{G}_D}{c^2} \int \frac{\epsilon_0 E^2 r^{\prime 2}\left(\sin ^3 \theta\right) d r^{\prime} d \theta}{\left.\left.\left|\left(r^2+r^{\prime 2}-2 r r^{\prime} \cos \theta\right)+\right| \vec{z}\right|^2\right|^{\frac{1+\delta}{2}}}
\times f_e(z) d^\delta z^{\prime}, \\
&&\lambda_{1, s}=  -\chi_s-2 \pi \frac{\hat{G}_D}{c^2} \int \frac{\epsilon_0 E^2 r^{\prime 2}\left(\cos ^2 \theta \sin \theta\right) d r^{\prime} d \theta}{\left.\left.\left|\left(r^2+r^{\prime 2}-2 r r^{\prime} \cos \theta\right)+\right| \vec{z}\right|^2\right|^{\frac{1+\delta}{2}}} 
\times f_e(z) d^\delta z^{\prime}.
\end{eqnarray}
Notice, that the explicit determination of the functions $\phi_{s}$ and $\chi_{s}$ is obtained when we analyze the internal structure of the brane and observe how the fields are located \cite{dahia2018rydberg}. Such localization is done through topological defects, which are structures capable of locating fermions. In Ref \cite{rubakov1983}  it is possible to observe that the Dirac fields are trapped through a Yukawa-type interaction. This is known as a thick brane with a delta-like location is then replaced by a regular wave function with a width of $\sigma$ in the transverse directions. However, the thin brane limit, where we neglect the thickness of the brane, we can say that a zero-width brane is a good approximation for states with higher quantum numbers. In this way, we can replace the $f_{m}(z)$ function with a delta-type function in function (\ref{TT7}) for an outer region. The gravitational potential for short distances created by the mass of the nucleus is
\begin{equation}\label{TT10}
\varphi_s=-\hat{G}_d \frac{M}{r^{1+\delta}},
\end{equation}
where $M$ is the nucleus mass. In our scenario, we can neglect the effects of the gravitational field $\chi_{s}$ created by electromagnetic energy since we are only taking relativistic corrections of order $1/c^{2}$ for the non-relativistic limit of the Dirac equation. Indeed, according to the metric (\ref{TT6}) this term is of the order of $1/c^{4}$. Furthermore, the gravitational dipole moment term is of the order of $1/c$ in (\ref{FW10}) which would result in a relativistic correction of the order of $1/c^{5}$ which makes its value negligible even considering the influence of the extra dimensions. The gravitational field $\chi(x)$ is of great relevance when investigating the energy of the system for the Rydberg states. For more details, see reference \cite{dahia2018rydberg}.

\section{Spin and Gravity In Brane}
\label{sgb}
In this section, we will investigate the gravitational effects on the non-relativistic Dirac Hamiltonian in the braneworld scenario. The starting point is, of course, the geometry of the system. We consider that on length scales above the thickness of the brane, the fields do not couple directly to the geometry of the volume of the supplementary space, thus their dynamics are governed by gravity through the geometry of the brane. This geometry must be isometric with respect to the geometry of the supplementary space, which allows us to take a coordinate system such that $z=0$ in (\ref{TT6}). Furthermore any deviations in the metric components within the brane are insignificant when compared to its thickness $\sigma$ \cite{dahia2018rydberg}. 

Since we neglect the effects of the gravitational field created by the function $\chi_{s}$ the gravitational field of the system is reduced to the gravitational field created by the mass of the nucleus. So we consider the metric (\ref{TT6}) in isotropic coordinates becomes
\begin{equation}\label{SG1}
ds^{2}=\left[1+\frac{2}{c^2} \varphi_s\right]c^{2}dt^{2}+\left[1-\frac{2}{c^2\left(1+\delta\right)} \varphi_s\right]\delta ijdx^{i}dx^{j} .
\end{equation}
By comparing (\ref{SG1}) with (\ref{SSS1}) at the weak field limit, we have that 
\begin{equation}\label{SG2}
v\approx1+\frac{1}{c^2}\varphi_{s}, \quad w\approx1-\frac{1}{c^2(1+\delta)}\varphi_{s}.
\end{equation}
This leads us to gradients (\ref{FW9})
\begin{equation}\label{SG3}
\vec{\phi}=-\frac{1+\delta}{c^2}\vec{g}_{d}, \quad \vec{f}=-\frac{2+\delta}{c^2}\vec{g}_{d}.
\end{equation}
Here, $\vec{g}_{d}$ is the gravitational field produced by the nucleus given as
\begin{equation}\label{SG4}
\vec{g}_{d}=-\tilde{G}_{d} M\frac{\vec{r}}{r^{(2+\delta)}}.
\end{equation}
Note that, if we consider ordinary spacetime, the metric (\ref{SG1}) becomes a Schwarzschild type metric, and the Newtonian gravitational field is recovered from (\ref{SG4}).

Now, by using equations (\ref{SG1}) to (\ref{SG4}) the Hamiltonian (\ref{FW10}) can be written as
\begin{equation}\label{SG5}
\mathcal{H}=\beta m c^2+\beta \frac{\vec{p}^2}{2 m}+\beta m\varphi_{s}-\beta\frac{\hbar^{2}}{4mc^{2}}(2+\delta)\vec{\nabla}\cdot\vec{g}_{d}-\beta\frac{\hbar}{2mc^{2}}(2+\delta)\vec{\Sigma}\cdot\left(\vec{g}_{d}\times\vec{p}\right)-\frac{\hbar(1+\delta)}{2c}\vec{\Sigma}\cdot\vec{g}_{d}.
\end{equation}
The first two terms define the free part of the Hamiltonian, the rest of energy and the kinetic term, respectively. The remaining terms describe the interaction with the gravitational field produced by the nucleus. In particular, the fourth term is known as the Darwin term responsible for the gravitational Zitterbewegung motion \cite{jentschura2013} and the fifth term the gravitational spin-orbit coupling term.

The main discussion of this paper is the analysis of gravitational momentum, the last term of the Hamiltonian (\ref{SG5}). This term preserves CP symmetry, but not C and P separately. In fact, notice that it does not appear in the Hamiltonian coupled with the matrix $\beta$ so it remains with the same sign for particles and antiparticles. This fact is related to the exact Foldy-Whoutuysen transformation \cite{Obukhov:2000ih}. A similar analysis arises when we analyze its parity. Furthermore, it must violate the equivalence principle since it results in a displacement between the center of mass and the center of gravity of the particle. However, averaging over the spins restores the equivalence principle on a macroscopic scale. In this way, the validity of the Dirac equation for fermions is demonstrated. A proposal for a possible experiment would be to observe the change in the equilibrium angle of the magnet when its magnetization is destroyed by heating above the Curie point \cite{Peres:1977yh}. The spin of the particle would precess about the vertical axis with a specific frequency, which, although faster than the prediction of general relativity, would still be too slow to be observed in neutron interference experiments. This is due to the fact that the interaction energy is very small, on the order of $\hbar g/c\approx\times10^{-23}\ \text{eV}$. 

In the brane scenario, the effects of the extra dimensions should significantly amplify the interaction energy for the gravitational dipole momentum. In fact, we observe that the coupling constant $k=(1+\delta)/2$ assumes different values for each number of extra dimensions, where for $\delta=0$ we recover the value $k=1/2$ as proposed in \cite{Obukhov:2000ih}. Furthermore, the gravitational field should be amplified by the additional dimensions as proposed by the ADD model \cite{Arkani-Hamed:1998jmv}. We can rewrite it in terms of the gravitational field (\ref{SG4}) and the gravitational constant (\ref{CG}) 
\begin{equation}\label{SG6}
H_{gm}=E(\delta)\vec{\Sigma}\cdot\vec{r}.   
\end{equation}
The energy $E(\delta)$, defined in terms of the number of extra dimensions $\delta$ and the compaction radius $R$. 
\begin{equation}\label{SG7}
E(\delta)=\frac{2(1+\delta)}{(2+\delta)}\frac{\Gamma\Big(\frac{3+\delta}{2}\Big)}{\pi^{(1-\delta)/2}}\frac{\hbar GM}{c}\frac{(2R)^{\delta}}{r^{2+\delta}}.
\end{equation}

At this point, we consider our system as a Hydrogen atom type system. In this way, the gravitational field is that created by the proton’s mass $M=1.672622\times10^{-27}\ \text{kg}$, and the distance  $r$  is the Bohr radius $a_{0}=5.2917721\times 10^{-11}\ \text{m}$. For $\delta=0$ we have the ordinary four-dimensional spacetime and the interaction energy for the gravitational momentum is
$
E(0)=\frac{\hbar}{2c}\frac{GM}{r^{2}}\approx10^{-41}\text{eV}.
$
This energy value is tiny when compared to the result in \cite{Obukhov:2000ih}, due to the fact that we are considering the gravitational field created by the proton. However, if we consider an extradimensional spacetime, from (\ref{SG7}) the values for energy of the gravitational momentum on the brane directly feel, the effects of the extra dimensions of the supplementary space can be significantly amplified. According to the ADD model, each extra compact dimension has its own compactification radius, so there must be different energy values for each extra dimension for equation (\ref{TT11}). The table below shows us explicitly the values of these for each extra dimension as well as its compaction radius.
\begin{table}[h!]
\centering
\begin{tabular}{||c c c ||} 
 \hline
 Extra Dimensions & Compaction Radius & Energy \\ [0.5ex] 
 \hline\hline
 $\delta=1$ & $\sim 10^{13}$ m & $\sim 10^{-17}$ eV \\ 
 $\delta=2$ & $\sim 10^{-6}$ m & $\sim 10^{-25}$ eV \\
 $\delta=3$ & $\sim 10^{-9}$ m & $\sim 10^{-35}$ eV\\
 $\delta=4$ & $\sim 10^{-11}$ m & $\sim 10^{-38}$ eV  \\ [1ex] 
 \hline
\end{tabular}
\caption{Estimation of the energy of the gravitational momentum term assuming different values for extra dimensions and different compaction radius according to the brane ADD model.}
\label{table 1}
\end{table}

In the above table, it is possible to observe that the energy values for the term gravitational dipole moment are significantly amplified by the additional dimensions. According to our analysis for a Universe with only one extra dimension $\delta=1$ the energy for the term gravitational dipole moment is of the order of $E(1)\sim10^{-17}\ \text{eV}$, that is, a single additional dimension amplifies the energy value by an order of $\sim10^{24}\ \text{eV}$ in relation to the four-dimensional ordinary value. However, since we are considering the ADD model, it proposes the existence of at least two extra dimensions $\delta\geq2$ since for a Universe with only one extra dimension its compaction radius would be $R\sim10^{13}\ \text{m}$ and, therefore, the dimension would already have been observed. Furthermore, only when $\delta=2$ do we have a compaction radius on the sub-millimeter scale $R\sim10^{-3}\ \text{m}$. This value is compatible with empirical data and deviations from the Newtonian law of gravitation are predicted \cite{Arkani-Hamed:1998jmv,arkani2000,arkani1999,gabadadze}.

Specifically, if $\delta=2$ the corresponding energy value for the gravitational dipole moment for a Universe with two extra dimensions is according to our analysis, $\sim10^{-25}\ \text{eV}$, which is a very small limit. However, it is clear that the extra dimension amplifies the energy value in relation to the energy of four-dimensional space-time by a factor of the order of $\sim10^{16}\ \text{eV}$. Although this value is far from experimental observations, it shows that theories of extra dimensions are extremely important from a theoretical point of view for high-energy physics \cite{salumbides2015,wan2015}. Thus, our analysis shows that deviations from the inverse square law for gravitation can lead us to spectroscopic constraints for the extra dimensions. It is expected that with the advent of interferometric techniques in atomic spectroscopy, the determination of extra spatial dimensions in our Universe will be feasible.

\section{Final Remarks}
\label{conc}
We investigate the Dirac theory for a fermion coupled to the gravitational field in the brane scenario proposed by the ADD model. In this model with large extra dimensions, gravity is affected by the extra space. In general, models with extra dimensions predict that the three-dimensional Newtonian interaction compared to gravitational interaction, can be amplified at short distances. In this way, we consider the effects of extra dimensions in the atomic domain to investigate the influence of the higher-dimensional gravitational field produced by the nucleus of a Hydrogen atom.

The formalism for the Dirac theory in curved spacetime in the braneworld scenario was presented, where it was possible to determine the non-relativistic Hamiltonian by means of an exact Foldy-Whoutuysen transformation. This exact method provides the Hamiltonian with a gravitational dipole moment contribution that is configured in an interaction between the electron spin and the gravitational field created by the proton that defines the atomic nucleus. This contribution, behaves differently for particles and antiparticles, and thus, violating the C and P symmetries separately. And also, violating the equivalence principle which is restored by averaging over the spins. In ordinary spacetime, the energy contribution of this term to the total energy of the system is tiny, on the order of $\sim 10^{-41}\ \text{eV}$. However, if we consider the braneworld scenario, the interaction energy for the gravitational dipole moment feels the effects of the extra dimensions of the supplementary space, so that its energy contribution is amplified.

Our analysis has fixed the spacetime with two extra dimensions, from which the ADD model becomes viable. In this scenario, the energy for the gravitational dipole moment obtained is $\sim10^{-25}\ \text{eV}$ (amplified of $\sim10^{16}$ order of magnitude in comparison to the energy for ordinary space-time). Although this result is still too small to be verified experimentally, it confirms the theoretical prediction about the influence of extra dimensions in atomic spectroscopy in the sense of amplifying the value of the energy of the terms affected by these dimensions (see again in Refs. \cite{salumbides2015,wan2015}). Furthermore, it suggests that in the future, with the advent of interferometry techniques, it will be feasible to detect traces of extra dimensions in atomic spectroscopy.

{\acknowledgments} EBM thanks to the Graduate Program in Physics at the Federal University of Campina Grande PPGF. 
MAA and EP acknowledge support from CNPq (Grant nos. 306398/2021-4, 304290/2020-3) and Paraiba State Research Foundation, Grant no. 0015/2019.

\end{document}